\documentclass[twocolumn,grl]{agu2001}
\usepackage{graphicx}
\authorrunninghead{DAUB ET AL.}

\titlerunninghead{SHEAR STRAIN LOCALIZATION}

\authoraddr{E. G. Daub, M. L. Manning, and J. M. Carlson
Physics Department, Broida Hall, University of
California, Santa Barbara, CA 93106, USA.
(edaub@physics.ucsb.edu)}

\begin{document}

%
%
%
\setkeys{Gin}{draft=false}
%
%

%
%

\title{Shear strain localization in elastodynamic rupture simulations}
%

%
%


\author{Eric G. Daub, M. Lisa Manning, and Jean M. Carlson}
\affil{Physics Department, University of California, Santa Barbara, CA 93106, USA}

\begin{abstract}
We study strain localization as an enhanced velocity weakening mechanism on
earthquake faults. Fault friction is modeled using Shear Transformation Zone
(STZ) Theory, a microscopic physical model for non-affine rearrangements in
granular fault gouge. STZ Theory is implemented in spring slider and dynamic rupture
models of faults. We compare dynamic shear localization to 
deformation that is uniform throughout the gouge layer, and find that 
localized slip enhances the velocity weakening of the gouge. Localized elastodynamic
ruptures have larger stress drops and higher peak slip rates than ruptures
with homogeneous strain.
\end{abstract}

%
%

%

\begin{article}

%
%

\section{Introduction}

The earthquake problem spans a vast range of length and time scales, from
contacts between individual grains up through tectonic networks of faults. 
One of the primary modeling challenges involves identifying the relevant physical
instabilities, and accurately and efficiently propagating this information
between scales.
In this
study, we investigate gouge-scale strain localization as a physical mechanism for
enhanced velocity weakening and apply it to the larger scale problem of
dynamic rupture propagation on faults. Strain localization has been observed
in a variety of contexts, including numerical simulations
of gouge [{\it Morgan and Boettcher}, 1999], experimental studies of laboratory faults [{\it Marone}, 1998], and
field observations of faults [{\it Chester and Chester}, 1998].

The formation of shear bands is a dynamic instability that
we resolve numerically by incorporating a gouge layer of finite width. This approach
differs from the common practice of implementing a slip weakening or rate and state
friction law on a planar fault. In our model, the dynamic behavior of the friction law determines
how strain is distributed throughout the gouge layer.
Within the gouge,
the material is governed by Shear Transformation Zone (STZ) Theory, a microscopic
physical model for non-affine deformation in amorphous materials [{\it Falk and Langer}, 1998, 2000].
We study the formation of localized shear bands, in contrast with homogeneous deformation.
For homogeneous deformation, strain is uniform throughout the gouge layer.
In shear bands, slip spontaneously localizes to an interface which is narrow even on the
scale of the gouge.
The dynamic instability associated with localization enhances the velocity weakening that occurs with homogeneous
deformation.
We find that the enhanced weakening due to shear
strain localization decreases the dynamic sliding friction
and increases the peak slip rates in dynamic earthquake models.
The results illustrate that gouge-scale strain localization
has fault-scale consequences.

\section{STZ Friction Law}

STZ Theory is a continuum model for amorphous materials, 
suitable for
predicting the larger scale constitutive behavior of a layer of fault gouge.
STZ Theory captures certain features of a wide array of amorphous materials, including 
deformation of metallic glasses [{\it Falk and Langer}, 1998, 2000], boundary lubrication [{\it Lemaitre and Carlson}, 2004], granular flow [{\it Lois et al.}, 2005], shear bands in glassy systems [{\it Manning
et al.}, 2008], and frictional weakening in dynamic earthquake rupture [{\it Daub and Carlson}, 2008].

\begin{figure}
\includegraphics[width=39pc]{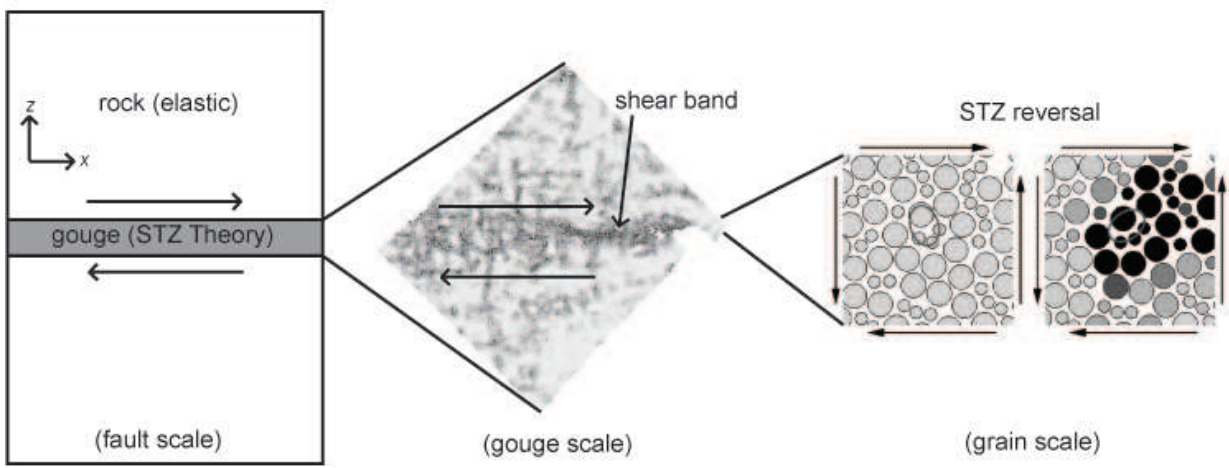}
\caption{\label{fig:stz}
An illustration of the range of scales in the earthquake rupture problem.
The scale moves to progressively smaller phenomena from left to right.
Left: fault scale model, with a thin layer of gouge described by STZ theory
sheared between elastic rocks. Center: close up of deformation inside the
gouge, where shear strain develops into a localized shear band (dark regions). 
Shear band image taken from {\it Falk and Shi} [2002], and re-oriented to match
the sense of shear of the fault and grains. 
Right: microscopic picture of
the grain scale, with an STZ undergoing 
transformation from a ``+'' oriented zone (left) to a ``$-$'' oriented zone
(right). As the gouge deforms plastically,
the ellipse drawn through the particles flips its orientation.
STZ diagram taken from {\it Falk and Langer} [1998].}
\end{figure}

A schematic of the fault is shown at the left in Fig.~\ref{fig:stz}. A layer of fault
gouge, which is modeled using STZ Theory, is sheared between elastic rocks. 
The plastic strain rate $\dot{\gamma}_{pl}$ can vary in the $z$-direction in the gouge (center
picture in Fig.~\ref{fig:stz}), 
though the shear stress $\tau$ is assumed to be uniform in the layer.

Non-affine rearrangements in the gouge occur in localized regions, called
shear transformation zones (STZs). STZs are modeled as bistable
zones that switch between two orientations under shear stress. 
An STZ can only flip once in the direction of the applied shear stress, so
STZs are continually created and annihilated to sustain plastic flow.
A schematic of an STZ switching orientation is shown at the right in Fig.~\ref{fig:stz}.

The plastic strain rate is influenced by two properties of the gouge:
the number of STZs, and
how these STZs change orientation. The number of STZs follows a Boltzmann distribution, with
an effective disorder temperature $\chi$ [{\it Langer}, 2004], and the STZs also change orientation
due to the applied stress.
These two contributions can be summarized as:
\begin{equation}
\label{eq:basicstzs}
\dot{\gamma}_{pl}=\exp\left(-1/\chi\right)R\left(\tau\right).
\end{equation}
The function $R(\tau)$ describes how the rate at which STZs change orientation
depends on the applied shear stress. The physics behind our choice for $R(\tau)$ is 
an Eyring model [{\it Eyring}, 1936], which matches the logarithimic
velocity dependence of laboratory faults. However, the enhancement of velocity
weakening that strain localization produces is independent of the details of $R(\tau)$.

Regions with higher effective temperature have more STZs, and undergo more
plastic strain. The effective temperature is distinct from the thermal temperature, but is similar to
the free volume [{\it Langer}, 2007], which was used as the state variable 
in previous STZ friction models [{\it Lemaitre}, 2002]. However, we expect the effective
temperature to exhibit dynamic behavior similar to thermal temperature, and we 
therefore include shear heating and diffusion terms in its governing partial 
differential equation. The friction equations and the parameters for our simulations can all be
found in Table~\ref{table:stz}. The details of the derivation can be found in {\it Manning et al.} [2007]
and {\it Lemaitre} [2002].

\begin{table}
\caption{\label{table:stz} Equations and parameters for the STZ constitutive law. The equations include the
specific version of Eq.~(\ref{eq:basicstzs}) and the partial differential equation governing
the effective temperature.}
\begin{tabular}{|c|c|}
\hline
\multicolumn{2}{|c|}{Equations} \\
\hline
\multicolumn{2}{|c|}{$\dot{\gamma}_{pl}=\exp\left(-1/\chi\right)2\epsilon/t_0\exp\left(-f_0\right)\left(1-\tau_y/\tau\right)\cosh\left(\tau/\sigma_d\right)$} \\
\multicolumn{2}{|c|}{$\partial \chi/\partial t=\dot{\gamma}_{pl}\tau/(c_0\tau_y)\left(1-\chi \log(\dot{\gamma}_0/\dot{\gamma}_{pl})/\chi_w\right)+\dot{\gamma}_{pl}D\partial^2 \chi/\partial z^2$} \\
\hline
Parameter & \multicolumn{1}{|c|}{Description} \\
\hline
$\epsilon= 10$ & \multicolumn{1}{|c|}{Typical number of particles in an STZ} \\
$t_0=10^{-6}$ s & \multicolumn{1}{|c|}{STZ rearrangement time scale} \\
$f_0= 118$ & \multicolumn{1}{|c|}{STZ rearrangement activation energy} \\
$\tau_y=50$ MPa & \multicolumn{1}{|c|}{Yield stress (below $\tau_y$, $\dot{\gamma}_{pl}=0$)} \\
$\sigma_d = 0.5$ MPa & \multicolumn{1}{|c|}{Stress fluctuation for STZ rearrangement} \\
$c_0=1$ & \multicolumn{1}{|c|}{Effective temperature specific heat} \\
$\dot{\gamma}_0=80000$ ${\rm s}^{-1}$ & \multicolumn{1}{|c|}{Strain rate at which $\chi$ diverges} \\
$\chi_w=0.5$ & \multicolumn{1}{|c|}{Stress weakens with strain rate if $\chi_w<1$} \\
$D=10^{-5}$ ${\rm m}^2$ & \multicolumn{1}{|c|}{Diffusion constant} \\
\hline
\end{tabular}
\end{table}

\section{Spring Slider Model}

To investigate the dynamics of localization, 
we study the STZ friction law (Table~\ref{table:stz})
coupled to a non-inertial single degree of freedom elastic slider. A layer
of gouge of width $2w$ separates two blocks, one of which is pulled by a 
spring of stiffness $k$ at a velocity $V_0$. We numerically solve for the 
stress and the effective temperature in the gouge layer using
finite differences with an explicit two step time integration scheme.
Boundary conditions on the effective temperature are periodic, as
the effective temperature dynamics are not particularly sensitive to the boundary
conditions.
We drive the slider from rest to a seismic slip rate of $V_0=$~1~m/s with a spring of
stiffness $k=$~100000~MPa/m. This approximates the rapid loading and slip acceleration
that occurs during dynamic seismic slip.
We set the half width of the gouge layer as $w=$~1~m and start with an 
initial shear stress of $\tau(t=0)=$~70~MPa.

\begin{figure}
\includegraphics[width=20pc]{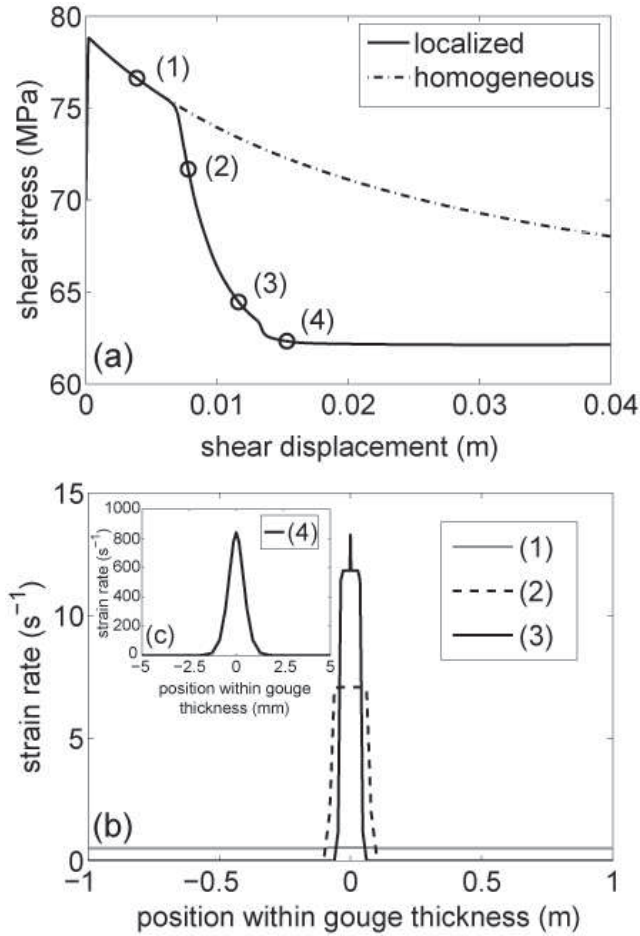}
\caption{\label{fig:slider}
(a) Plot of shear stress as a function of shear displacement for the spring slider model.
Comparison between dynamic strain localization and homogeneous deformation
reveals that localized strain exhibits more rapid weakening.
(b) Strain rate profiles for four representative
shear displacements during localized strain, indicated in plot (a). (1) An initial period of broad deformation occurs
before (2) strain dynamically localizes. (3) A narrower, diffusion-limited shear band develops with further shear. 
(c) Inset: The narrow shear band eventually accommodates all of the deformation.}
\end{figure}

A special case of the spring slider dynamics arises for homogeneous initial conditions for
the effective temperature. In this case, by symmetry all subsequent deformation is homogeneous.
We set the initial effective temperature to be uniformly $\chi(t=0)=0.018$.
A plot of stress as a function of displacement
is illustrated in Fig.~\ref{fig:slider}, which shows that
stress weakens with displacement in a manner
similar to the laboratory-based Dieterich-Ruina friction law 
[{\it Dieterich}, 1979], and the
STZ Free Volume law considered by {\it Daub and Carlson} [2008].

In general, we expect the initial conditions for effective temperature in the gouge to
reflect the inherent heterogeneity of fault zones [{\it Chester and Chester}, 1998]. 
A localized shear band dynamically forms in this case, due to the dynamic response
of the friction equations.
A shear band, with a width that scales with the diffusion constant $\sqrt{D}$, develops and ultimately accomodates all of the strain in the material.

While a narrow shear band results from any non-uniform initial conditions,
we focus on an idealized scenario to illustrate the dynamic evolution
of the shear band. We add a small amplitude ($\Delta\chi=2\times10^{-7}$)
symmetric step perturbation in the center of the gouge layer 
of half width 0.1~m to the otherwise uniform initial effective temperature ($\chi(t=0)=0.018$).

We compare the shear stress as a function of displacement
for both dynamically localized strain and homogeneous strain
in Fig.~\ref{fig:slider}(a). Uniform strain requires
about 0.3~m of displacement for the shear stress to stabilize to $\tau=$~63.38~MPa.
Dynamic localization drops the sliding stress to $\tau=$~62.14~MPa
much more rapidly, and
illustrates that localization enhances the velocity
weakening of the gouge.

Figures~\ref{fig:slider}(b) and \ref{fig:slider}(c) show snapshots of the strain rate in the 
gouge layer for dynamically localizing strain at a series
of representative points along the stress versus displacement curve.
During the earliest stages
of displacement, strain occurs nearly uniformly throughout the
gouge layer (plot (1) in Fig.~\ref{fig:slider}(b)). The
effect of the initial perturbation is negligible, and the strain rate is uniform.
The displacement before localized strain begins depends on the magnitude of
the perturbation in the initial effective temperature. This is because a region
with an elevated effective temperature also has a higher strain rate 
(Eq.~(\ref{eq:basicstzs})). The shear heating term in the effective temperature
equation (Table~\ref{table:stz}) is proportional to the strain rate, so regions of increased effective
temperature heat more rapidly. Therefore, for larger initial perturbations,
the local effective temperature increases more rapidly, and less 
displacement is needed before the strain dynamically localizes.

The strain rate profile once localization begins is shown in curve (2) in Fig.~\ref{fig:slider}(b).
The dynamic instability in the friction law causes strain to localize
to a shear band, and friction weakens rapidly. 
A narrower shear band (plots (3) in Fig.~\ref{fig:slider}(b) and (4) in 
Fig.~\ref{fig:slider}(c)) develops with further displacement.
By symmetry, the narrow shear band occurs in the center of the gouge, and
its width is determined by the diffusion constant.
Once the strain has completely localized to this diffusion-limited shear band,
the frictional strength stabilizes. While the shear heating
and diffusion terms do not exactly cancel each other, there is no noticeable change in the
effective temperature on the time scale of an earthquake rupture and the gouge slides stably.

\section{Dynamic Ruptures}

In this section, we investigate the impact of strain localization on
the propagation of spontaneous elastodynamic ruptures. We model the fault gouge
as a thin layer with half-width $w$ between two homogeneous, isotropic, 
linear elastic solids. We solve for the elastodynamic
response for 2D anti-plane slip using a spectral boundary integral method 
[{\it Perrin et al.}, 1995]. The
elastodynamic equation is solved simultaneously with the friction law (Table~\ref{table:stz}).

The spectral boundary integral method can only accomodate explicit time steps.
The diffusion term in the effective temperature equation imposes a time step restriction that makes implementation
in a rupture code impractical. However, ignoring diffusion in the spring slider model
leads to nearly the same stress response as a function of displacement.
Therefore, as a first effort to incorporate localization
into dynamic rupture models, we omit the diffusion term in the effective temperature
evolution equation.

We consider a simple fault 2~km in length along strike, with a gouge half width of 1~m. 
The friction parameters in 
Table~\ref{table:stz} are spatially uniform along strike and throughout the gouge width
(not including the effective temperature, which we solve for dynamically).
The shear stress on the fault is initially $\tau(t=0)=70$~MPa,
except for  
a small patch of width 0.2~km at the center of the fault where the
stress is 79~MPa to initiate rupture. 
We consider two different rupture scenarios, analogous to those in the spring slider section:
one where shear strain localizes dynamically, and one where
deformation is homogeneous. The initial effective
temperature does not vary along strike, and within the gouge
layer it is the same as for the corresponding spring slider model for each
rupture scenario.

\begin{figure}
\includegraphics[width=20pc]{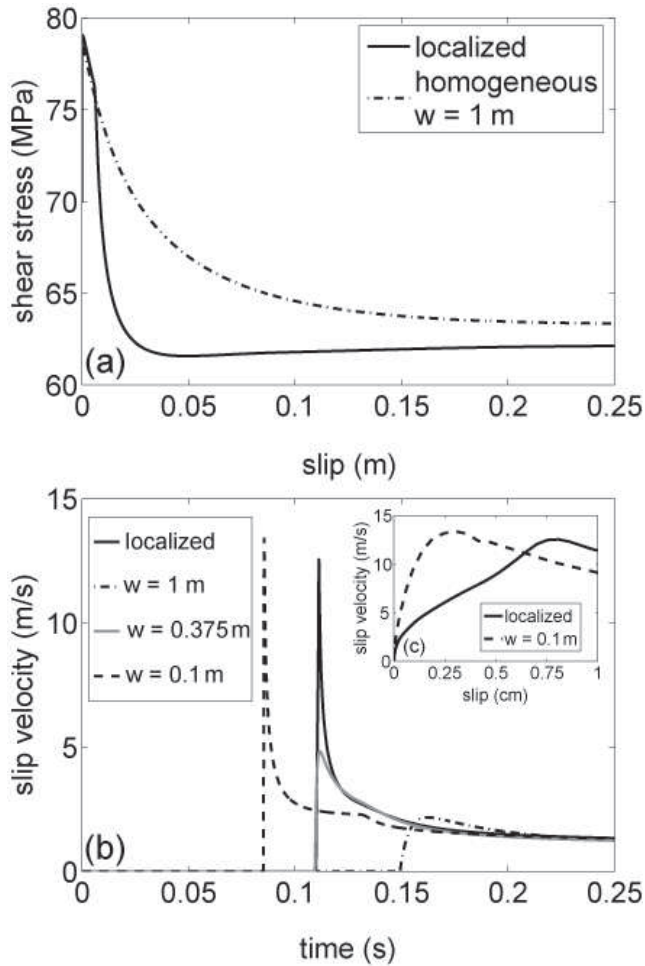}
\caption{\label{fig:rupture}
Dynamic rupture evolution at a point 0.35~km from the hypocenter.
(a) Comparison of shear stress as a function of slip. Dynamic localization of deformation produces more rapid velocity weakening
than the rupture with homogeneous strain.
(b) Plot of slip rate as a function of time. The dynamic strain localization rupture is compared with
a host of models with homogeneous strain. None of the values of the 
imposed gouge width $w$ can match both the peak slip rate and rupture front arrival of the
rupture with localized strain.
(c) Inset: Slip rate as a function of slip for the localized and narrowest width 
homogeneous rupture. The more rapid acceleration of slip in the narrowest homogeneous rupture
is distinct from the localized model.}
\end{figure}

A plot comparing how stress weakens with slip for the two ruptures
is shown in Fig.~\ref{fig:rupture}(a). During the initial stages
of slip, the curves are indistinguishable. For the later stages of slip, shear stress weakens much
more rapidly due to the dynamic instability of 
localization, increasing the stress drop. 
Homogeneous deformation and dynamic localization produce very
different slip rates, as can be seen in Fig.~\ref{fig:rupture}(b).
The rupture front arrives earlier and has a higher peak slip rate when
strain localization occurs.

To illustrate the importance of the dynamic instability, we contrast
our results for localized shear with two additional homogeneous ruptures of
different fixed gouge widths $w$ chosen to match particular aspects of the localized rupture.
Slip velocity as a function of time is plotted at a point
0.35~km from the hypocenter for all models in Fig.~\ref{fig:rupture}(b).
The properties that we compare are the peak slip velocity and
the time at which slip initiates.
The rupture front in the intermediate model ($w=$~0.375~m) matches the
arrival time of the localized rupture, but the peak slip rate is smaller. 
For the narrowest gouge thickness ($w=$~0.1~m), 
we see peak slip rates similar to the localized
rupture but earlier arrival.
Figure~\ref{fig:rupture}(c) plots the slip rate as a function of slip for 
the homogeneous rupture with $w=$~0.1~m and the localized rupture. 
This clearly shows that the initial broad deformation in 
the localized rupture does not simply delay the rupture, but also lessens
the slip acceleration during the earliest stages of slip.

\section{Discussion}

The main effect of dynamic strain localization in our simulations is to
provide a mechanism for enhanced velocity weakening.
The dynamic instability
that produces localization has two effects on the shear stress: a
reduced dynamic sliding stress, and a very rapid decrease in stress with 
slip.
Allowing for enhanced weakening makes our faults more unstable --
both the stress drop and the peak slip rates increase for dynamic localization
relative to homogeneous slip.

The width of the narrow shear band in our simulations is determined by the diffusion length scale 
$\sqrt{D}$. For different driving rates, other shear band
widths are possible.

Other weakening mechanisms have been discussed in the context of
rapid seismic slip, including pore fluid pressurization [{\it Lachenbruch}, 1980],
flash heating at asperity contacts [{\it Tullis and Goldsby}, 2003], 
production of a thixotropic silica gel [{\it Di Toro et al.}, 2004],
and frictional melting [{\it Di Toro et al.}, 2006]. Most of these
are due to (regular) thermal effects. Localization complements these, as 
localization occurs before significant
change in the thermal temperature. Rupture models that employ both could examine
how thermal weakening and shear localization interact.

The enhanced weakening effect of dynamic localization occurs regardless of 
the choice of the function $R(\tau)$ in Eq.~(\ref{eq:basicstzs}).
Other forms of $R(\tau)$
will still exhibit more rapid weakening
during localized deformation. This indicates that shear localization can affect a
variety of materials, and have consequences for other geophysical systems.

\begin{acknowledgments}
The authors thank Eric Dunham for providing
the dynamic rupture code used in this study.
This work was supported by the James S. McDonnell Foundation, the David and
Lucile Packard Foundation, NSF grant number DMR-0606092, and the NSF/USGS 
Southern California Earthquake Center, funded by NSF Cooperative Agreement
EAR-0106924 and USGS Cooperative Agreement 02HQAG0008. 
\end{acknowledgments}

%
%
%
%
%
%
%
%


\end{article}

%
%
%

\end{document}